# Evidence for Magnon-Assisted Exciton Radiation in a Layered Antiferromagnetic Semiconductor

Yingchen Peng[#,1] Yanan Ge[#,2] Wentao Chen,[1] Zihan Wang,[3] Kang Wang,[1*] Kezhao Du,[3] Xingzhi Wang,[2] Ye Yang[1†]

[1]State Key Laboratory of Physical Chemistry of Solid Surfaces, College of Chemistry and Chemical Engineering, Xiamen University, Xiamen 361005, P.R. China

[2]Department of Physics, Xiamen University, Xiamen 361005, P.R. China;

[3]Fujian Provincial Key Laboratory of Advanced Materials Oriented Chemical Engineering, Fujian Normal University, Fuzhou 350007, P. R. China;

[#]These authors contributed equally.

*Contact author: kangwang@xmu.edu.cn; †Contact author: ye.yang@xmu.edu.cn

**Abstract.** Layered van der Waals (vdW) magnetic semiconductors open a new avenue for exploring intertwined excitonic and magnetic phenomena. Here, we investigate this interplay in the vdW antiferromagnet $MnPS_3$, uncovering an exceptionally long exciton lifetime (~100 $\mu s$) below the Néel temperature ($T_N$). We demonstrate that the exciton lifetime is governed by phonon-assisted nonradiative recombination and thus exhibits a strong temperature dependence. In contrast, the exciton radiative recombination shows a distinct temperature dependence that is sensitive to magnetic order. Below $T_N$, the temperature dependence of the radiative recombination rate is consistent with a magnon-assisted emission pathway, while above $T_N$ it reflects the combined effects of short-range spin correlations and phonons. These findings not only establish $MnPS_3$ as a compelling candidate for excitonic devices due to its long-lifetime and correlation with magnetic orders but also provide crucial insights into the interplay between excitons, spins, and lattice in vdW magnetic semiconductors.

Layered van der Waals (vdW) magnetic semiconductors have sparked intense research interest due to their potential in ultrathin spintronic and magnonic applications.[1-3] These materials serves as an ideal platform for studying magnetism in two dimensions,[4-9] leading to the discoveries of many intriguing physical phenomena tied to the magnetism.[10-19] Among these, the interaction between magnetism and excitons is particularly compelling,[20-29] offering an optical knob to manipulate the magnetism on ultrafast time scales. However, implementation of these functionalities requires the magnetic excitons with sufficiently long lifetime, which demands a fundamental understanding of their recombination mechanisms. This is challenging due to the interwind coupling between exciton, lattice and spin degrees of freedom in these materials.[13,18,21,23,30-41] Despite their pivotal role, the exciton recombination mechanisms and their connections with magnetic ordering remain largely unexplored in vdW antiferromagnetic semiconductors.

Here, we investigated the ultrafast exciton dynamics and its correlation with magnetic orders in a vdW $MnPS_3$ antiferromagnet using a combination of transient absorbance (TA) and time-resolved photoluminescence (TRPL) spectroscopies. We observe that excitons form from free photocarriers on a sub-picosecond timescale across a wide temperature range, a process that is notably independent of magnetic order. In stark contrast to this rapid formation, excitons take nearly 100 $\mu s$ to recombine below the Néel temperature, leading to exceptionally long exciton lifetimes. The exciton lifetime exhibits a strong temperature dependence, which is quantitatively captured by a multi-phonon mediated nonradiative recombination model. Furthermore, by examining the temperature-dependent photoluminescence intensity and lifetime, we extract the temperature trend of the radiative recombination rate and find that, in the antiferromagnetic phase, the exciton radiative recombination proceeded through a magnon-assisted mechanism. Our work successfully disentangles the influences of magnons and phonons on the exction recombination process in a vdW antiferromagnet, demonstrating that they govern the radiative and nonradiative recombination pathways, respectively.

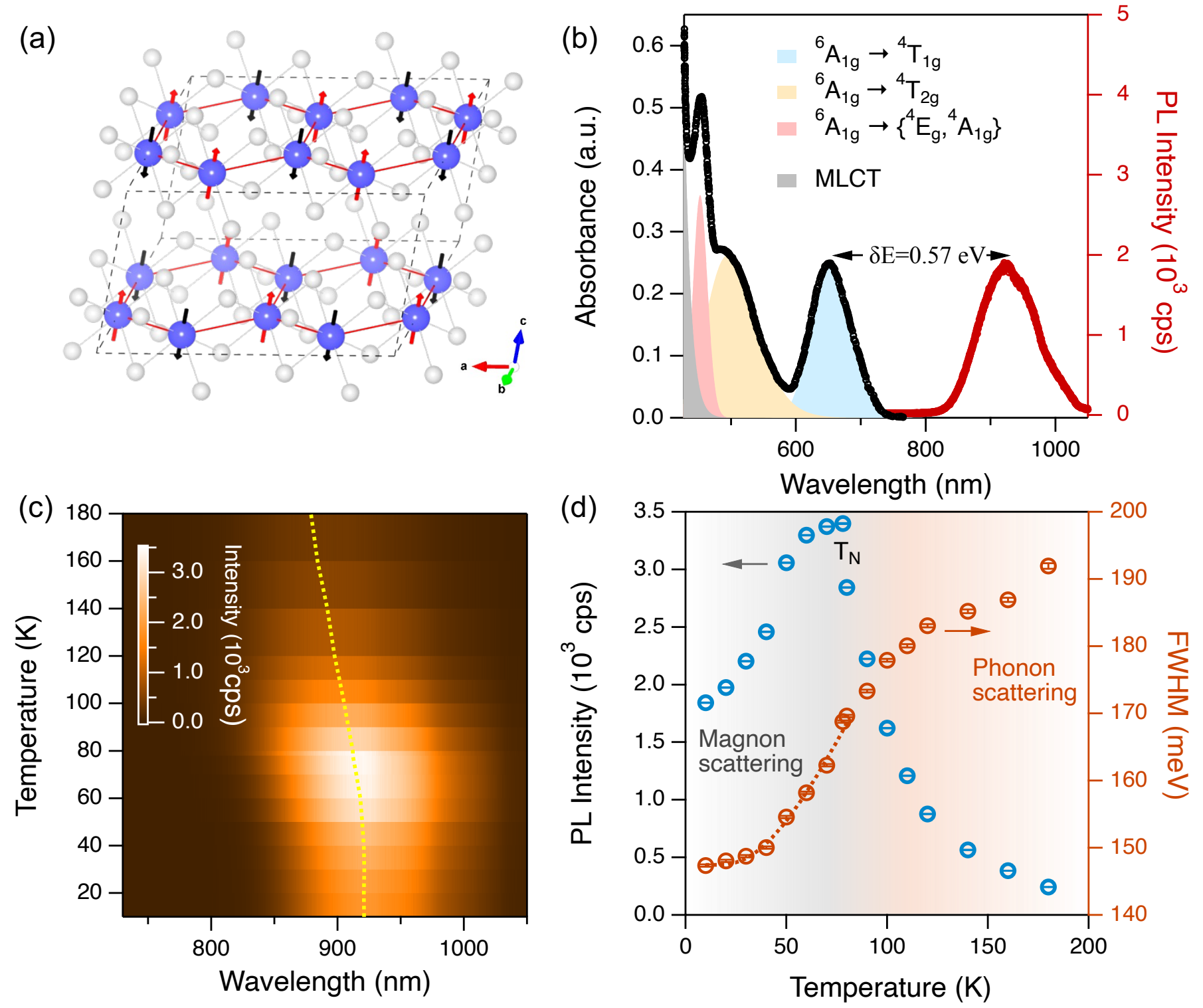

**Figure 1. Photoluminescence of excitons in $MnPS_3$.** (a) Schematic illustration of the lattice and spin structures of $MnPS_3$. (b) Absorbance (black curve) and photoluminescence (red curve) spectra measured at 10 K. The absorbance spectrum is adapted from ref. 44. (c) Pseudocolor image of the temperature-dependent photoluminescence spectra. The intensity is indicated by the color scale bar. The yellow dotted curve marks the peak position. (d) The photoluminescence peak intensities (blue circles) and full width at half maxima (orange circles) of the photoluminescence at various temperatures. The error bars are obtained from the uncertainty of Gaussian fitting. The dash curve represents the linewidth broadening due to magnon scattering.

**Correlation between exciton photoluminescence and magnetic order.** High-quality $MnPS_3$ crystals were grown using a chemical vapor transport method.[42] In these crystals, manganese ions ($Mn^{2+}$), octahedrally coordinated by sulfur ions, form a honeycomb lattice in the $ab$ planes, and these $Mn^{2+}$ ions are antiferromagnetically coupled with their nearest neighbors below the Néel temperature of $T_N$= 78 K (Fig. 1a).[43] The absorbance spectrum of $MnPS_3$ (Fig. 1b, black curve) shows weak *d-d* transition bands below the metal-to-ligand charge transfer (MLCT) gap, corresponding to the electric-dipole transitions from the $Mn^{2+}$ sextet ground state to four quartet states.[12,44] The transitions, though normally forbidden by both Laporte and spin selection rules, are rendered weakly allowed by symmetry breaking.[44,45] The photoluminescence spectrum (Fig. 1b, red curve) shows a broad peak with a large Stokes shift (0.57 eV), indicative of significant lattice relaxation in exciton state. The emission in strongly correlated antiferromagnets has been previously attributed to Frenkel excitonic states,[45-49] consistent with the theoretically predicted large exciton binding energy in $MnPS_3$.[39,50]

The temperature dependence of the photoluminescence spectra is visualized in Fig. 1c. As temperature rises, the photoluminescence peak shifts towards higher energy (Fig. 1c, dotted curve). This trend mirrors the temperature dependent shift of the lowest *d-d* transition (i.e., $^6A_{1g} \rightarrow {}^4T_{1g}$ ) but is opposite to the behavior of higher *d-d* transitions.[44,51] The photoluminescence peak intensity ($I_{PL}$) first increases with temperature, peaks precisely at $T_N$, and then gradually declines (Fig. 1d, blue circles). The temperature dependence of the photoluminescence peak intensity closely follows that of the integrated intensity (Fig. S1). This nonmonotonic trend has been widely reported in previous studies of $MnPS_3$,[49,52,53] and a similar correlation between the peak temperature of $I_{PL}$ and $T_N$ has also been observed in antiferromagnetic $Mn_xZn_{1-x}PS_3$ alloys with 0.6 < x < 1.[53] This coincidence has been interpreted as evidence for magneto-optical coupling in these antiferromagnets.[49,53] However, the nonmonotonic temperature dependence of $I_{PL}$ alone is not a unique signature of long-range antiferromagnetic order because it persists even in non-magnetic $Mn_xZn_{1-x}PS_3$ alloys (x < 0.5) where long-range antiferromagnetic order is completely suppressed.[53] This apparent contradiction arises probably because $I_{PL}$ is determined by the competition between radiative and nonradiative recombination processes, and the underlying recombination mechanisms will be discussed later. The photoluminescence linewidth (Fig. 1d, orange circles), quantified by the full-width at half-maximum (FWHM), is also linked to the magnetic order. Below $T_N$, the FWHM increases rapidly with temperature, but the slope is explicitly reduced above $T_N$, forming a distinct kink at the transition. At elevated temperatures (T > 150 K), the FWHM resumes a rapid increase, suggesting a crossover where the phonon scattering probably becomes the dominant broadening mechanism. The temperature dependence of $I_{PL}$ and FWHM observed here is consistent with prior literatures.[49,52,53] The temperature trend of FWHM below $T_N$ is reminiscent of the boson broadening model,[38,54] in which the linewidth broadening is proportional to the boson occupation number (i.e., $\propto [\exp(\hbar\omega/k_B T) - 1]^{-1}$). A fit to this model (Fig. 1d, dotted curve) yields a boson energy of $\hbar\omega$ = 12$\pm$1 meV, which coincides with magnon modes at the Brillouin zone boundary.[55,56] Thus, temperature dependent broadening in the spin ordered phase may arise from magnon scattering.

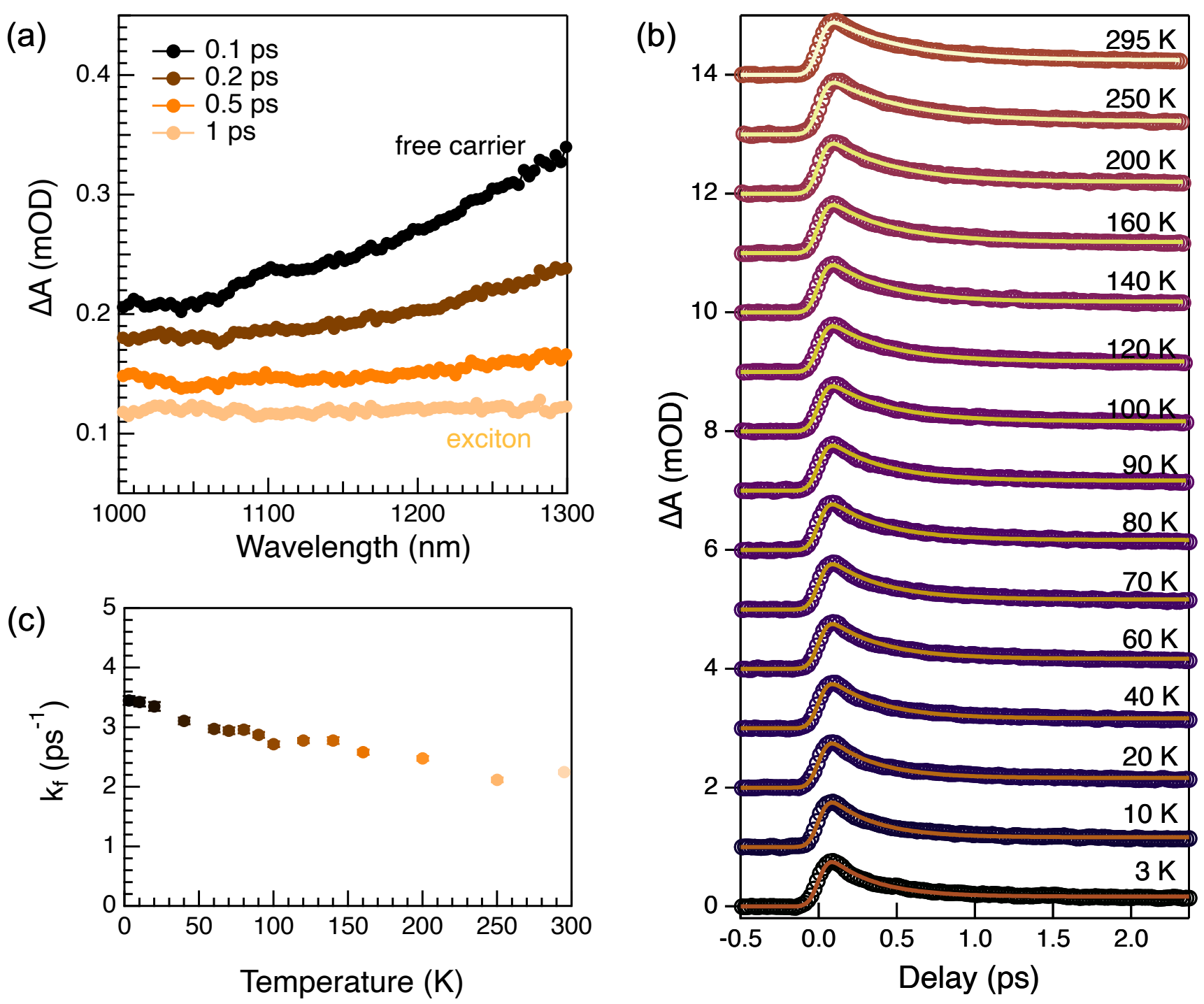

**Figure 2. Exciton formation dynamics.** (a) Temporal evolution of transient absorbance (TA) spectra measured at 3 K (b) Exciton formation dynamics represented by the TA spectral kinetics for different temperatures. The solid curves represent single-exponential fitting functions. (c) Temperature dependences of the exciton formation rate.

**Temperature dependence of exciton formation dynamics**. The correlation between excitons and magnetic order, revealed in the steady-state photoluminescence, motivates an investigation into the exciton formation dynamics at different temperatures using femtosecond transient absorbance (fs-TA) spectroscopy. The sample was pumped at photon energy of 3.1 eV (above the MLCT gap) to directly create free holes and electrons in their respective bands. The resulting photoinduced absorbance change ($\Delta A$) was probed in the near-infrared region, a spectral window where probe photon energy is significantly smaller than optical transitions from the ground state to any excited states, thus $\Delta A$ must be associated with transitions between different excited states. As shown in Fig. 2a, at a short delay (e.g., 0.1 ps), the magnitude of $\Delta A$ increases with wavelength, characteristic of the free carrier intraband absorption.[57,58] With increasing delay, the signal size of $\Delta A$ quickly decreases, and, simultaneously, its upward-trend spectral shape progressively flattens. The temporal evolution of the spectral shape is more clearly shown in Fig. S2, which closely resembles that associated with self-trapped exciton (STE) formation in transition metal oxides and vdW magnetic materials.[34,35,57,59] At long delays, spectral shape remains unchanged, whose decay kinetics matches the exciton recombination dynamics measured by the time-resolved photoluminescence (demonstrated later), confirming that the residual TA signal at long delays should be associated with the optical transitions from the excitonic state to higher energy levels.

To probe the exciton formation mechanism, we performed TA measurements at different temperatures. The resulting $\Delta A$ kinetics features a fast decay component (Fig. 2b), which corresponds to the sub-picosecond spectral evolution from the free carrier signal to excitonic feature. This component directly captures the exciton formation process. A single-exponential fit to these kinetics yields the exciton formation rate ($k_f$), which decreases monotonically with increasing temperature (Fig. 2c), consistent with STE formation in transition metal oxides.[60] Despite the signature of self-trapping revealed by the TA spectra and kinetics, a definitive identification of STEs would require direct evidence of exciton-induced local lattice distortion, which is beyond the scope of the present work. Notably, the temperature trend of the exciton formation rate shows no discontinuity or abrupt change near $T_N$, indicating that the exciton formation process is independent of the spin ordering.

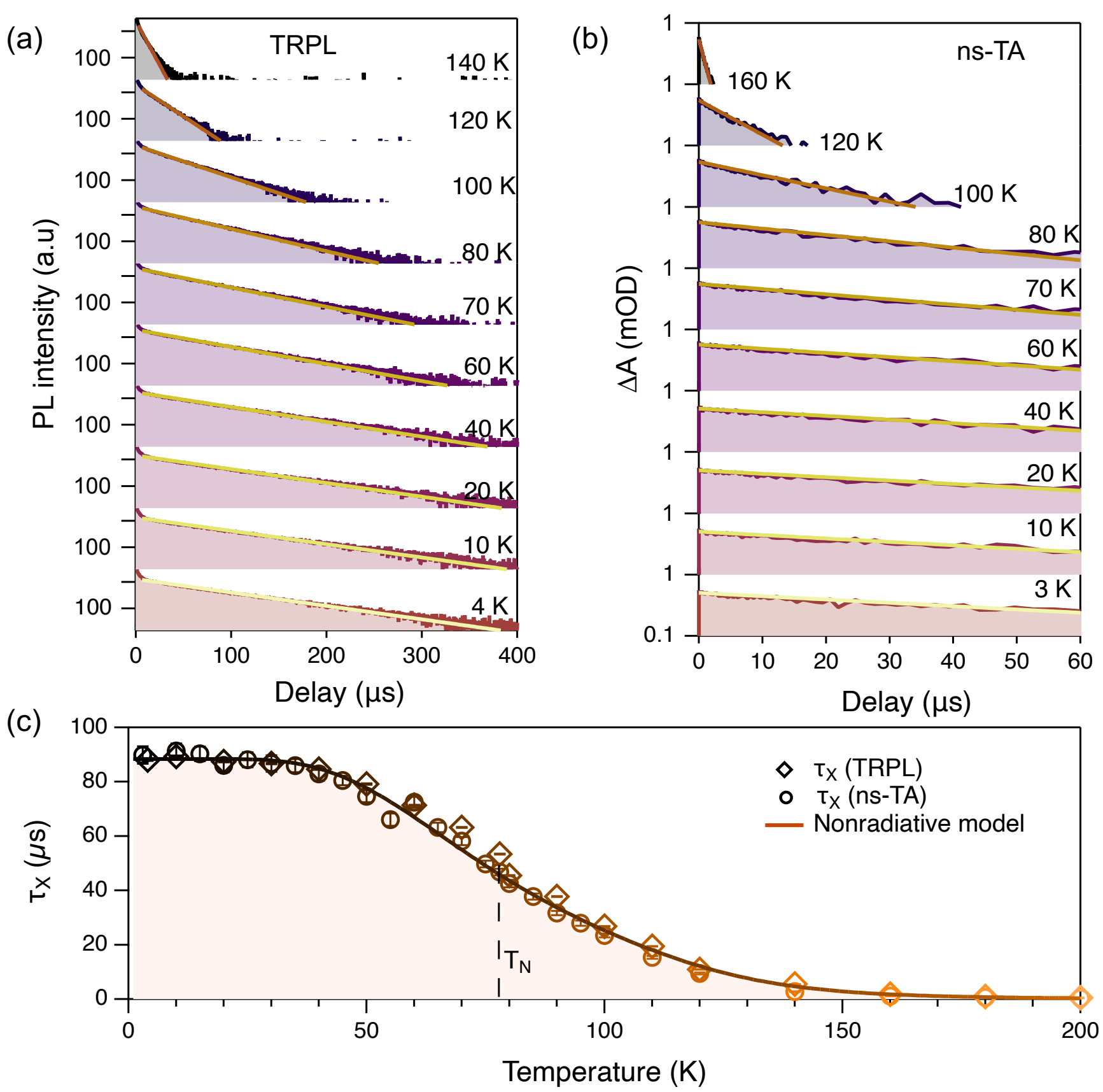

**Figure 3. Temperature-dependent exciton lifetimes.** Temperature dependent exciton recombination dynamics measured by (a) time-resolved photoluminescence (TRPL) and (b) nanosecond transient absorbance (ns-TA). The vertical axes are plot in logarithmic scale. (c) The exciton lifetimes determined from TRPL and ns-TA at different temperatures. The solid curve represents the exciton lifetime predicted by the nonradiative recombination model.

**Temperature dependence of exciton lifetime**. We measured exciton recombination dynamics at different temperatures using two complementary techniques: time-resolved photoluminescence (TRPL) and nanosecond transient absorbance (ns-TA) with temporal resolutions of 13 $\mu s$ and 1 $ns$, respectively. Fig. 3a and 3b present the TRPL and ns-TA kinetics at representative temperatures. The TRPL kinetics exhibit a single exponential decay at all temperatures (Fig. 3a, solid curves), indicative of a monomolecular recombination mechanism. The TRPL decays slowly below $T_N$, signifying long exciton lifetimes, and the decay rate accelerates rapidly as temperature increases above $T_N$. The ns-TA measurements probe the long-lived excitonic feature identified in the fs-TA spectra (Fig. 3b). Despite the different delay ranges, the decay rates of the ns-TA kinetics are nearly identical to those of the TRPL kinetics at corresponding temperatures. The spectral assignment of the excitonic TA feature discussed afore was made based on this consistence. The ns-TA measurement also reveals that the exciton lifetime shortens by over four orders of magnitude when the temperature is raised from cryogenic to room temperature (Fig. S3). Note that the sub-picosecond exciton formation cannot be resolved in the ns-TA measurements due to the limited temporal resolution.

The exciton lifetime ($\tau_X$) extracted from both methods shows excellent agreement, which increases monotonically as temperature drops and eventually saturates at the lower temperatures (Fig. 3c). $\tau_X$ is determined by the sum of the radiative ($k_r$) and nonradiative ($k_{nr}$) recombination rate, $\tau_X = (k_r + k_{nr})^{-1}$, and the photoluminescence quantum yield ($\phi_{PL}$) is given by the ratio of $\frac{k_r}{k_r+k_{nr}}$. Given the very low photoluminescence quantum yield, $k_r$ should be much smaller than $k_{nr}$, leading to an approximated relation of $\tau_X \sim {k_{nr}}^{-1}$. According to the nonradiative recombination model for the strong exciton-phonon coupling case, $k_{nr}$ can be described by the following relation [61-66], ${k_{nr}}^{-1} \propto \exp\left[\frac{E_a}{\hbar\omega\left(n+\frac{1}{2}\right)}\right]$, where $\hbar\omega$ is the energy of the coupled phonon mode, $n$ is the phonon occupation number and $E_a$ is activation energy, defined as the energy barrier from bottom of the exciton potential energy surface to the intersection of the ground and excitonic potential energy surfaces. We find that this model successfully captures the temperature trend of $k_{nr}$ ($\tau_X$), i.e., $k_{nr}$ ($\tau_X$) decreases (increases) quickly with decreasing temperature due to the reduction in $n$, but it saturates at low temperatures (where $n \rightarrow 0$, the recombination at this regime primarily proceeds via nuclear quantum tunneling).[67,68] A fit of this model to the experimental $\tau_X$ data (Fig. 3c, solid curve) yields the phonon energy of $\hbar\omega$ = 20$\pm$1 meV. This value corresponds to an $A_g$ phonon mode with frequency of ~150 $cm^{-1}$, which has previously reported to be entangled with magnons.[33,69,70] Note that the exciton lifetime is primarily governed by nonradiative recombination channels, which are mediated by phonons.

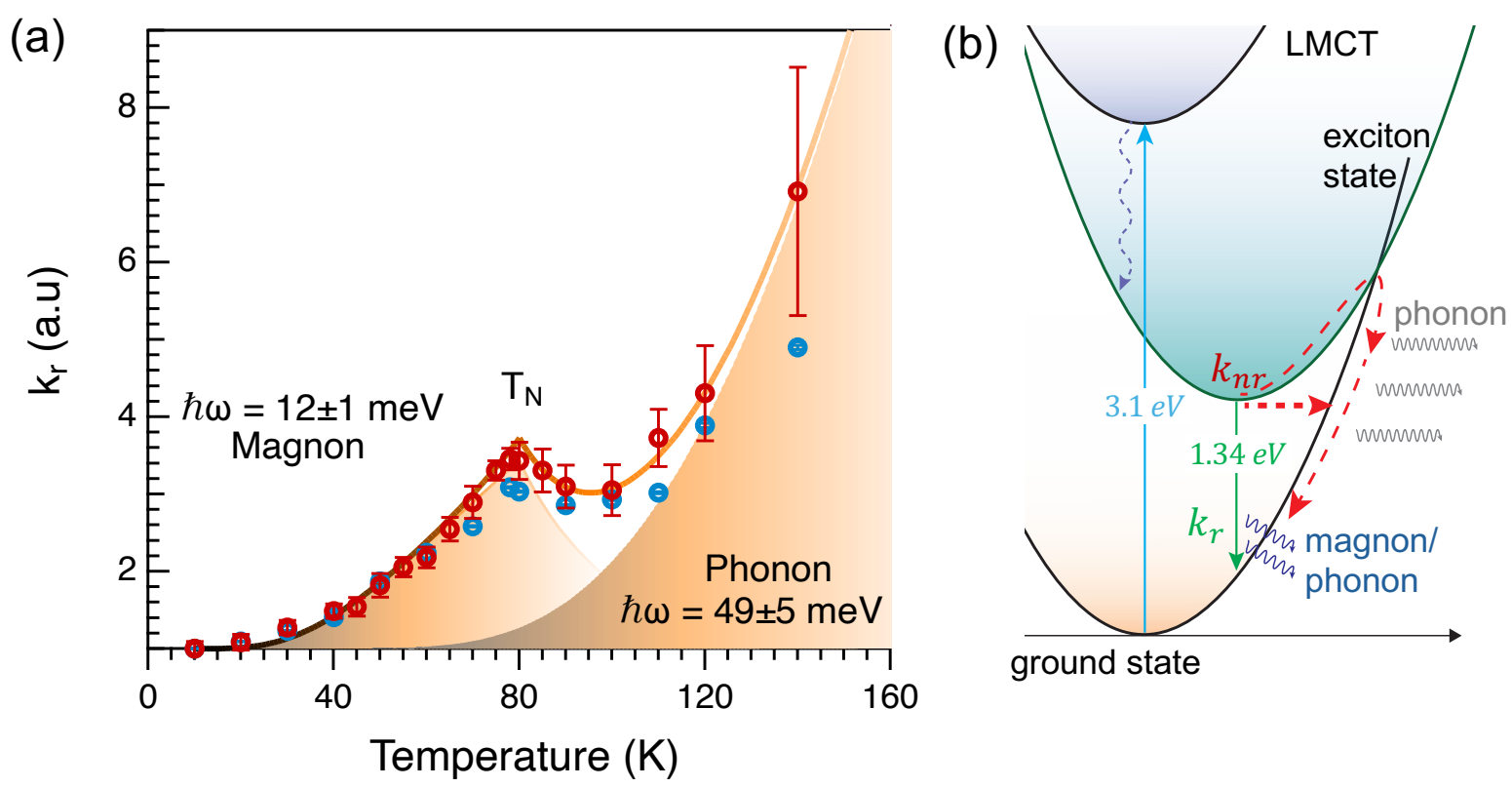

**Figure 4. Temperature trend of radiative exciton recombination.** (a) Temperature trend of $k_r$ is extracted from exciton lifetime and photoluminescence intensity (see main text) and normalized at 10 K. The error bars are obtained by propagating the uncertainties of the exciton lifetime and PL intensity. The solid curve represents the Boson assisted radiative recombination model, and the shades indicate the respective contributions from magnons and phonons. (b) Schematic illustration of the exciton recombination mechanisms in $MnPS_3$. The red dash arrows represent nonradiative recombination channels, including nuclear quantum tunneling and thermal activation. The cyan arrow represents the magnon assisted radiative recombination.

Under identical excitation conditions, $I_{PL}$ is proportional to $\phi_{PL}$, and thus $I_{PL} \propto \frac{k_r}{k_r+k_{nr}} = k_r\tau_X$. It is worth noting that the temperature trends for $\tau_X$ and $I_{PL}$ diverge below $T_N$ (Fig. S4), indicating that $k_r$ must have a temperature dependence opposite to that of $k_{nr}$. Although absolute values of $k_r$ cannot be directly measured, the temperature trend of $k_r$ can be extracted from the experimentally measured $I_{PL}$ and $\tau_X$ at different temperatures because $k_r \propto I_{PL}{\tau_X}^{-1}$ (blue and red circles, Fig. 4a). The temperature dependence of $k_r$ can be consistently reproduced using different samples (Fig. S5). $k_r$ exhibits a steep increase as temperature rises below $T_N$ and then experiences a decrease precisely at $T_N$, which is followed by resumption of another rapid increase. The temperature trend of $k_r$ below $T_N$ is well described by a boson-assisted radiative exciton recombination model in which the emissive transition probability is proportional to $e^{-\hbar\omega/k_BT}$, where $\hbar\omega$ is the boson energy.[71] A fit to this model (Fig.4a, solid curve below $T_N$) yields a boson energy of $\hbar\omega = 12 \pm 1$ meV, consistent with the energy of magnons at the Brillouin zone boundary.[55,56] We note that the energy of a $B_g$ optical phonon mode at the zone-boundary, within experimental uncertainty, also overlap with the extracted bosonic energy.[72,73] Therefore, based on energy considerations alone, phonon contributions to the radiative recombination cannot be excluded.

Importantly, the distinction between phonon- and magnon-assisted mechanisms can be revealed by the temperature dependence of $k_r$ . A purely phonon-assisted radiative recombination process would lead to a monotonic increase of $k_r$ with temperature, reflecting the increasing phonon population, and would not give rise to a reduction across $T_N$. This behavior is supported by measurements on a $Mn_{0.42}Zn_{0.58}PS_3$ alloy (Fig. S6), in which long-range antiferromagnetic order is suppressed within the concerned temperature range.[53,73] In this case, $k_r$ increases monotonically over the entire measured temperature range (Fig. S7). In contrast, $k_r$ in $MnPS_3$ exhibits a clear reduction upon crossing $T_N$ (Fig. 4a), coinciding with the loss of long-range magnetic order (i.e., the disappearance of propagating magnons). Furthermore, the exciton emission in $MnPS_3$ originates from a spin-forbidden transition. In this context, coupling to magnons provides a natural pathway to relax spin selection rules, as magnons carry spin angular momentum and can directly participate in the recombination process. By contrast, phonons do not carry spin and can only assist indirectly through lattice modulation and spin-orbit-mediated state mixing. These considerations suggest that magnons play an important role in the radiative recombination process below $T_N$, although mixed contributions cannot be fully excluded.

Just above $T_N$, the modest decrease in $k_r$, rather than a dramatic drop due to the loss of propagating magnons, implies the presence of an additional emission pathway. Although the spin fluctuations destroy the long-range spin ordering above $T_N$ , the spin correlation does not instantly vanish, and correlation length progressively decays as temperature increases.[74-76] In this regime, the long-range ordering is replaced by the correlated spin clusters with the average size scaling with the correlation length. We therefore propose a phenomenological hypothesis wherein the effectiveness of the coupling between excitons and spin waves is commensurate with the spin cluster size because a larger region of coherent spin order provides a more extended interaction volume for the exciton, potentially enhancing the coupling strength, and then $k_r$ may scale with the correlation length. This phenomenological assumption is introduced to account for the observed modest transition for $k_r$ across the phase change from long-range to short-range magnetic order.

An exciton radiative recombination model that incorporates the combined effects of magnons, spin clusters and phonons can reproduce the temperature trend of $k_r$ throughout the temperature range (Fig. 4a, solid curve). The energy of the participating phonons is determined to be 49$\pm$5 meV, corresponding to an $A_g$ phonon mode at 385 $cm^{-1}$.[77] This phonon-mediated contribution mainly accounts for the $k_r$ growth trend at temperatures well above $T_N$ in the paramagnetic phase. The characteristic phonon energy extracted here is different from that extracted from the non-radiative recombination (Fig. 3c), indicating that radiative and non-radiative processes couple to different phonon modes. The relatively large fitting uncertainty of the phonon energy arises from the limited temperature range above $T_N$ (it is difficult to record reliable $I_{PL}$ at higher temperature due to the low quantum yield). While our phenomenological model captures the temperature trend of $k_r$, a more rigorous theoretical framework is required to fully elucidate the underlying mechanisms, particularly the role of short-range spin-ordered structures.

The exciton recombination mechanisms in antiferromagnetic $MnPS_3$ uncovered in this work are schematically illustrated in Fig. 4b. Optical excitation at LMCT band directly generates free carriers (or free excitons), which rapidly collapse into the localized Frenkel excitons with significant lattice relaxation. These Frenkel excitons return to the ground state via two competing pathways, radiative and nonradiative recombination. The probability of a direct radiative transition is inherently low because it is forbidden by both Laporte (parity) and spin selection rules. This limitation is circumvented with the assistance of magnons, a mechanism we confirm through the distinct temperature dependence of $k_r$. Despite this magnon-assisted pathway, the exciton lifetime remains primarily governed by $k_{nr}$, evidenced by the low photoluminescence quantum yield. At sufficiently low temperatures, where the coupled phonons are frozen out, the thermal activation pathway is suppressed, allowing the nuclear quantum tunneling channel to dominate $k_r$, accounting for the saturation of the exciton lifetime at this regime (Fig. 3c). In the paramagnetic phase (above $T_N$) the situation evolves. Nonradiative recombination continues to be dominated by phonon-mediated pathways. In contrast, the radiative mechanism becomes more complex, influenced by a combination of short-range magnetic fluctuations and phonons. Our findings provide a microscopic understanding of the correlation between photoluminescence intensity and long-range magnetic order. The exceptionally long exciton lifetime, coupled with the strong exciton-magnon interaction in antiferromagnetic $MnPS_3$, highlights the potential of this material for developing novel magneto-optical applications.

**Acknowledgement:** Y.Y. acknowledges the National Natural Science Foundation of China under grant no. 22473093 and 22461160285, Fundamental Research Funds for the Central Universities under grant nos. 20720240150 and 20720220011. X.W. acknowledges the support of the National Key R&D Program of China (2022YFA1602704), the National Natural Science Foundation of China (62275225), the Fundamental Research Funds for the Central Universities under grant nos. 20720220034 and the 111 Project (B16029). The authors thank Dr. Xue Lou from the Instrumentation and Service Center for Molecular Sciences at Westlake University.

**References:**

[1] A. V. Chumak, V. I. Vasyuchka, A. A. Serga, and B. Hillebrands, Magnon spintronics, Nat. Phys. **11**, 453 (2015).

[2] X.-X. Zhang, L. Li, D. Weber, J. Goldberger, K. F. Mak, and J. Shan, Gate-tunable spin waves in antiferromagnetic atomic bilayers, Nat. Mater. **19**, 838 (2020).

[3] S. Rahman, J. F. Torres, A. R. Khan, and Y. Lu, Recent Developments in van der Waals Antiferromagnetic 2D Materials: Synthesis, Characterization, and Device Implementation, ACS Nano **15**, 17175 (2021).

[4] K. S. Burch, D. Mandrus, and J.-G. Park, Magnetism in two-dimensional van der Waals materials, Nature **563**, 47 (2018).

[5] H. Chu, C. J. Roh, J. O. Island, C. Li, S. Lee, J. Chen, J.-G. Park, A. F. Young, J. S. Lee, and D. Hsieh, Linear Magnetoelectric Phase in Ultrathin MnPS3 Probed by Optical Second Harmonic Generation, Phys. Rev. Lett. **124**, 027601 (2020).

[6] K. Kim, S. Y. Lim, J.-U. Lee, S. Lee, T. Y. Kim, K. Park, G. S. Jeon, C.-H. Park, J.-G. Park, and H. Cheong, Suppression of magnetic ordering in XXZ-type antiferromagnetic monolayer NiPS3, Nat. Commun. **10**, 345 (2019).

[7] J.-U. Lee, S. Lee, J. H. Ryoo, S. Kang, T. Y. Kim, P. Kim, C.-H. Park, J.-G. Park, and H. Cheong, Ising-Type Magnetic Ordering in Atomically Thin FePS3, Nano Lett. **16**, 7433 (2016).

[8] Q. Zhang, K. Hwangbo, C. Wang, Q. Jiang, J.-H. Chu, H. Wen, D. Xiao, and X. Xu, Observation of Giant Optical Linear Dichroism in a Zigzag Antiferromagnet FePS3, Nano Lett. **21**, 6938 (2021).

[9] X.-X. Zhang, S. Jiang, J. Lee, C. Lee, K. F. Mak, and J. Shan, Spin Dynamics Slowdown near the Antiferromagnetic Critical Point in Atomically Thin FePS3, Nano Lett. **21**, 5045 (2021).

[10] A. Zong, Q. Zhang, F. Zhou, Y. Su, K. Hwangbo, X. Shen *et al.*, Spin-mediated shear oscillators in a van der Waals antiferromagnet, Nature **620**, 988 (2023).

[11] X. Wang, J. Cao, H. Li, Z. Lu, A. Cohen, A. Haldar, H. Kitadai, Q. Tan, K. S. Burch, D. Smirnov, W. Xu, S. Sharifzadeh, L. Liang, and X. Ling, Electronic Raman scattering in the 2D antiferromagnet NiPS3, Sci. Adv. **8**, eabl7707 (2022).
[12] J.-Y. Shan, M. Ye, H. Chu, S. Lee, J.-G. Park, L. Balents, and D. Hsieh, Giant modulation of optical nonlinearity by Floquet engineering, Nature **600**, 235 (2021).
[13] E. Ergeçen, B. Ilyas, D. Mao, H. C. Po, M. B. Yilmaz, J. Kim, J.-G. Park, T. Senthil, and N. Gedik, Magnetically brightened dark electron-phonon bound states in a van der Waals antiferromagnet, Nat. Commun. **13**, 98 (2022).
[14] S. Y. Kim, T. Y. Kim, L. J. Sandilands, S. Sinn, M.-C. Lee, J. Son *et al.*, Charge-Spin Correlation in van der Waals Antiferromagnet NiPS3, Phys. Rev. Lett. **120**, 136402 (2018).
[15] C.-T. Kuo, M. Neumann, K. Balamurugan, H. J. Park, S. Kang, H. W. Shiu, J. H. Kang, B. H. Hong, M. Han, T. W. Noh, and J.-G. Park, Exfoliation and Raman Spectroscopic Fingerprint of Few-Layer NiPS3 Van der Waals Crystals, Sci. Rep. **6**, 20904 (2016).
[16] E. Ergeçen, B. Ilyas, J. Kim, J. Park, M. B. Yilmaz, T. Luo, D. Xiao, S. Okamoto, J.-G. Park, and N. Gedik, Coherent detection of hidden spin–lattice coupling in a van der Waals antiferromagnet, Proc. Natl Acad. Sci. USA **120**, e2208968120 (2023).
[17] D. Afanasiev, J. R. Hortensius, M. Matthiesen, S. Mañas-Valero, M. Šiškins, M. Lee, E. Lesne, H. S. J. van der Zant, P. G. Steeneken, B. A. Ivanov, E. Coronado, and A. D. Caviglia, Controlling the anisotropy of a van der Waals antiferromagnet with light, Sci. Adv. **7**, eabf3096 (2021).
[18] S. Liu, A. Granados del Águila, D. Bhowmick, C. K. Gan, T. Thu Ha Do, M. A. Prosnikov, D. Sedmidubský, Z. Sofer, P. C. M. Christianen, P. Sengupta, and Q. Xiong, Direct Observation of Magnon-Phonon Strong Coupling in Two-Dimensional Antiferromagnet at High Magnetic Fields, Phys. Rev. Lett. **127**, 097401 (2021).
[19] D. Jana, P. Kapuscinski, I. Mohelsky, D. Vaclavkova, I. Breslavetz, M. Orlita, C. Faugeras, and M. Potemski, Magnon gap excitations and spin-entangled optical transition in the van der Waals antiferromagnet NiPS3, Phys. Rev. B **108**, 115149 (2023).
[20] S. Kang, K. Kim, B. H. Kim, J. Kim, K. I. Sim, J.-U. Lee *et al.*, Coherent many-body exciton in van der Waals antiferromagnet NiPS3, Nature **583**, 785 (2020).
[21] Y. J. Bae, J. Wang, A. Scheie, J. Xu, D. G. Chica, G. M. Diederich *et al.*, Exciton-coupled coherent magnons in a 2D semiconductor, Nature **609**, 282 (2022).
[22] G. M. Diederich, J. Cenker, Y. Ren, J. Fonseca, D. G. Chica, Y. J. Bae, X. Zhu, X. Roy, T. Cao, D. Xiao, and X. Xu, Tunable interaction between excitons and hybridized magnons in a layered semiconductor, Nat. Nanotechnol. **18**, 23 (2023).
[23] K. Hwangbo, Q. Zhang, Q. Jiang, Y. Wang, J. Fonseca, C. Wang, G. M. Diederich, D. R. Gamelin, D. Xiao, J.-H. Chu, W. Yao, and X. Xu, Highly anisotropic excitons and multiple phonon bound states in a van der Waals antiferromagnetic insulator, Nat. Nanotechnol. **16**, 655 (2021).
[24] C. A. Belvin, E. Baldini, I. O. Ozel, D. Mao, H. C. Po, C. J. Allington, S. Son, B. H. Kim, J. Kim, I. Hwang, J. H. Kim, J.-G. Park, T. Senthil, and N. Gedik, Exciton-driven antiferromagnetic metal in a correlated van der Waals insulator, Nat. Commun. **12**, 4837 (2021).
[25] F. Dirnberger, R. Bushati, B. Datta, A. Kumar, A. H. MacDonald, E. Baldini, and V. M. Menon, Spin-correlated exciton–polaritons in a van der Waals magnet, Nat. Nanotechnol. **17**, 1060 (2022).
[26] D. S. Kim, D. Huang, C. Guo, K. Li, D. Rocca, F. Y. Gao *et al.*, Anisotropic Excitons Reveal Local Spin Chain Directions in a van der Waals Antiferromagnet, Adv. Mater. **35**, 2206585 (2023).

[27] Z. Wang, X.-X. Zhang, Y. Shiomi, T.-h. Arima, N. Nagaosa, Y. Tokura, and N. Ogawa, Exciton-magnon splitting in the van der Waals antiferromagnet MnPS3 unveiled by second-harmonic generation, Physical Review Research **5**, L042032 (2023).

[28] D. Wang, H. Chen, Y. Pang, X. Zou, and W. Duan, Band-spin-valley coupled exciton physics in antiferromagnetic MnPS3, Phys. Rev. B **112**, 075306 (2025).

[29] M. Onga, Y. Sugita, T. Ideue, Y. Nakagawa, R. Suzuki, Y. Motome, and Y. Iwasa, Antiferromagnet–Semiconductor Van Der Waals Heterostructures: Interlayer Interplay of Exciton with Magnetic Ordering, Nano Lett. **20**, 4625 (2020).

[30] G. M. Diederich, M. Nguyen, J. Cenker, J. Fonseca, S. Pumulo, Y. J. Bae, D. G. Chica, X. Roy, X. Zhu, D. Xiao, Y. Ren, and X. Xu, Exciton dressing by extreme nonlinear magnons in a layered semiconductor, Nat. Nanotechnol. **20**, 617 (2025).

[31] W. He, Y. Shen, K. Wohlfeld, J. Sears, J. Li, J. Pelliciari, M. Walicki, S. Johnston, E. Baldini, V. Bisogni, M. Mitrano, and M. P. M. Dean, Magnetically propagating Hund's exciton in van der Waals antiferromagnet NiPS3, Nat. Commun. **15**, 3496 (2024).

[32] B. Datta, P. C. Adak, S. Yu, A. Valiyaparambil Dharmapalan, S. J. Hall, A. Vakulenko *et al.*, Magnon-mediated exciton–exciton interaction in a van der Waals antiferromagnet, Nat. Mater. **24**, 1027 (2025).

[33] T. T. Mai, K. F. Garrity, A. McCreary, J. Argo, J. R. Simpson, V. Doan-Nguyen, R. V. Aguilar, and A. R. H. Walker, Magnon-phonon hybridization in 2D antiferromagnet MnPSe3, Sci. Adv. **7**, eabj3106 (2021).

[34] T. Yin, J.-Y. You, Y. Huang, H. T. Thu Do, M. A. Prosnikov, W. Zhao, M. Serra, P. C. M. Christianen, Z. Sofer, H. Sun, Y. P. Feng, and Q. Xiong, Signature of Ultrafast Formation and Annihilation of Polaronic States in a Layered Ferromagnet, Nano Lett. **22**, 7784 (2022).

[35] X. Li, A. Wang, H. Chen, W. Tao, Z. Chen, C. Zhang, Y. Li, Y. Zhang, H. Shang, Y.-X. Weng, J. Zhao, and H. Zhu, Ultrafast Spontaneous Localization of a Jahn-Teller Exciton Polaron in Two-Dimensional Semiconducting CrI3 by Symmetry Breaking, Nano Lett. **22**, 8755 (2022).

[36] Y. Li, G. Liang, C. Kong, B. Sun, and X. Zhang, Charge-Transfer-Mediated Exciton Dynamics in Van der Waals Antiferromagnet NiPS3, Adv. Funct. Mater. **34**, 2402161 (2024).

[37] A. Shcherbakov, K. Synnatschke, S. Bodnar, J. Zerhoch, L. Eyre, F. Rauh, M. W. Heindl, S. Liu, J. Konecny, I. D. Sharp, Z. Sofer, C. Backes, and F. Deschler, Solution-Processed NiPS3 Thin Films from Liquid Exfoliated Inks with Long-Lived Spin-Entangled Excitons, ACS Nano **17**, 10423 (2023).

[38] C.-H. Ho, T.-Y. Hsu, and L. C. Muhimmah, The band-edge excitons observed in few-layer NiPS3, npj 2D Mater Appl **5**, 8 (2021).

[39] M. Birowska, P. E. Faria Junior, J. Fabian, and J. Kunstmann, Large exciton binding energies in MnPS3 as a case study of a van der Waals layered magnet, Phys. Rev. B **103**, L121108 (2021).

[40] F. Mertens, D. Mönkebüscher, U. Parlak, C. Boix-Constant, S. Mañas-Valero, M. Matzer, R. Adhikari, A. Bonanni, E. Coronado, A. M. Kalashnikova, D. Bossini, and M. Cinchetti, Ultrafast Coherent THz Lattice Dynamics Coupled to Spins in the van der Waals Antiferromagnet FePS3, Adv. Mater. **35**, 2208355 (2023).

[41] J. E. Nitschke, L. Sternemann, M. Gutnikov, K. Schiller, E. Coronado, A. Omar, G. Zamborlini, C. Saraceno, M. Stupar, A. M. Ruiz, D. L. Esteras, J. J. Baldoví, F. Anders, and M. Cinchetti, Tracing the ultrafast buildup and decay of d-d transitions in FePS3, Newton **1**, 100019 (2025).

[42] K.-z. Du, X.-z. Wang, Y. Liu, P. Hu, M. I. B. Utama, C. K. Gan, Q. Xiong, and C. Kloc, Weak Van der Waals Stacking, Wide-Range Band Gap, and Raman Study on Ultrathin Layers of Metal Phosphorus Trichalcogenides, ACS Nano **10**, 1738 (2016).

[43] K. Okuda, K. Kurosawa, S. Saito, M. Honda, Z. Yu, and M. Date, Magnetic Properties of Layered Compound MnPS3, J. Phys. Soc. Jpn. **55**, 4456 (1986).

[44] V. Grasso, F. Neri, P. Perillo, L. Silipigni, and M. Piacentini, Optical-absorption spectra of crystal-field transitions in MnPS3 at low temperatures, Phys. Rev. B **44**, 11060 (1991).

[45] D. D. Sell, Review of Magnon‐Sideband Experiments, J. Appl. Phys. **39**, 1030 (1968).

[46] R. S. Meltzer, M. Y. Chen, D. S. McClure, and M. Lowe-Pariseau, Exciton-Magnon Bound State in MnF2 and the Exciton Dispersion in Mn2 and RbMnF3, Phys. Rev. Lett. **21**, 913 (1968).

[47] R. E. Dietz, A. E. Meixner, H. J. Guggenheim, and A. Misetich, Intrinsic luminescence of MnF2, J. Lumin. **1-2**, 279 (1970).

[48] D. D. Sell, R. L. Greene, and R. M. White, Optical Exciton-Magnon Absorption in MnF2, Phys. Rev. **158**, 489 (1967).

[49] Y. Zhou, K. He, H. Hu, G. Ouyang, C. Zhu, W. Wang *et al.*, Strong Neel Ordering and Luminescence Correlation in a Two-Dimensional Antiferromagnet, Laser Photonics Rev. **16**, 2100431 (2022).

[50] M. Rybak, P. E. Faria Junior, T. Woźniak, P. Scharoch, J. Fabian, and M. Birowska, Magneto-optical anisotropies of two-dimensional antiferromagnetic MnPX3 from first principles, Phys. Rev. B **109**, 054426 (2024).

[51] S. L. Gnatchenko, I. S. Kachur, V. G. Piryatinskaya, Y. M. Vysochanskii, and M. I. Gurzan, Exciton-magnon structure of the optical absorption spectrum of antiferromagnetic MnPS3, Low. Temp. Phys. **37**, 144 (2011).

[52] Y. Xing, H. Chen, A. Zhang, Q. Hao, M. Cai, W. Chen, L. Li, D. Peng, A. Yi, M. Huang, X. Wang, and J. Han, Optical Detection of Weak Magnetic Transitions via Photoluminescence in Cr-Doped Van Der Waals Antiferromagnets, Adv. Opt. Mater. **13**, 2402549 (2025).

[53] A. Harchol, S. Zuri, E. Ritov, F. Horani, M. Rybak, T. Woźniak, A. Eyal, Y. Amouyal, M. Birowska, and E. Lifshitz, Tuning magnetic and optical properties in MnxZn1−xPS3 single crystals by the alloying composition, 2D Mater. **11**, 035010 (2024).

[54] A. D. Wright, C. Verdi, R. L. Milot, G. E. Eperon, M. A. Pérez-Osorio, H. J. Snaith, F. Giustino, M. B. Johnston, and L. M. Herz, Electron–phonon coupling in hybrid lead halide perovskites, Nat. Commun. **7**, 11755 (2016).

[55] A. R. Wildes, S. Okamoto, and D. Xiao, Search for nonreciprocal magnons in MnPS3, Phys. Rev. B **103**, 024424 (2021).

[56] A. R. Wildes, B. Roessli, B. Lebech, and K. W. Godfrey, Spin waves and the critical behaviour of the magnetization in MnPS3, J. Phys.: Condens. Matter **10**, 6417 (1998).

[57] Y. Y. Fan, Y. M. Lin, K. Wang, K. H. L. Zhang, and Y. Yang, Intrinsic polaronic photocarrier dynamics in hematite, Phys. Rev. B **103**, 085206 (2021).

[58] P. Y. Yu and M. Cardona, *Fundamentals of Semiconductors* (Springer-Verlag Berlin Heidelberg, 2010), 4 edn., p.^pp. 305.

[59] J. Zhang, J. Shi, Y. Chen, K. H. L. Zhang, and Y. Yang, Bimolecular Self-Trapped Exciton Formation in Bismuth Vanadate, J Phys Chem Lett **13**, 9815 (2022).

[60] Y. Zhang, C. Zhang, X. Huang, Z. Yang, K. H. L. Zhang, and Y. Yang, Barrierless Self-Trapping of Photocarriers in Co3O4, J Phys Chem Lett **12**, 12033 (2021).

[61] J. J. Hopfield, Electron Transfer Between Biological Molecules by Thermally Activated Tunneling, Proc. Natl Acad. Sci. USA **71**, 3640 (1974).
[62] C. L. Xie and D. N. Hendrickson, Mechanism of spin-state interconversion in ferrous spin-crossover complexes: direct evidence for quantum mechanical tunneling, J. Am. Chem. Soc. **109**, 6981 (1987).
[63] C. H. Henry and D. V. Lang, Nonradiative capture and recombination by multiphonon emission in GaAs and GaP, Phys. Rev. B **15**, 989 (1977).
[64] H. Sumi, Multiphonon Nonradiative Recombination due to Successive Electron and Hole Capture by a Deep-Level Defect in Semiconductors, Phys. Rev. Lett. **47**, 1333 (1981).
[65] R. Englman and J. Jortner, The energy gap law for radiationless transitions in large molecules, Mol. Phys. **18**, 145 (1970).
[66] B. K. Ridley, Multiphonon, non-radiative transition rate for electrons in semiconductors and insulators, J. Phys. C: Solid State Phys. **11**, 2323 (1978).
[67] Y. Fan, Y. Lin, K. H. L. Zhang, and Y. Yang, Recombination of Polaronic Electron-Hole Pairs in Hematite Determined by Nuclear Quantum Tunneling, J Phys Chem Lett **12**, 4166 (2021).
[68] A. M. Stoneham, Non-radiative transitions in semiconductors, Rep. Prog. Phys. **44**, 1251 (1981).
[69] D. Vaclavkova, A. Delhomme, C. Faugeras, M. Potemski, A. Bogucki, J. Suffczyński, P. Kossacki, A. R. Wildes, B. Grémaud, and A. Saúl, Magnetoelastic interaction in the two-dimensional magnetic material MnPS3 studied by first principles calculations and Raman experiments, 2D Mater. **7**, 035030 (2020).
[70] K. Kim, S. Y. Lim, J. Kim, J.-U. Lee, S. Lee, P. Kim, K. Park, S. Son, C.-H. Park, J.-G. Park, and H. Cheong, Antiferromagnetic ordering in van der Waals 2D magnetic material MnPS3 probed by Raman spectroscopy, 2D Mater. **6**, 041001 (2019).
[71] B. Segall and G. D. Mahan, Phonon-Assisted Recombination of Free Excitons in Compound Semiconductors, Phys. Rev. **171**, 935 (1968).
[72] T. Babuka, M. Makowska-Janusik, A. V. Peschanskii, K. E. Glukhov, S. L. Gnatchenko, and Y. M. Vysochanskii, Electronic and vibrational properties of pure MnPS3 crystal: Theoretical and experimental investigation, Comput. Mater. Sci. **177**, 109592 (2020).
[73] R. Oliva, E. Ritov, F. Horani, I. Etxebarria, A. K. Budniak, Y. Amouyal, E. Lifshitz, and M. Guennou, Lattice dynamics and in-plane antiferromagnetism in MnxZn1-xPS3 across the entire composition range, Phys. Rev. B **107**, 104415 (2023).
[74] A. R. Wildes, H. M. Rønnow, B. Roessli, M. J. Harris, and K. W. Godfrey, Static and dynamic critical properties of the quasi-two-dimensional antiferromagnet MnPS3, Phys. Rev. B **74**, 094422 (2006).
[75] A. R. Wildes, S. J. Kennedy, and T. J. Hicks, True two-dimensional magnetic ordering in MnPS3, J. Phys.: Condens. Matter **6**, L335 (1994).
[76] Y.-J. Sun, Q.-H. Tan, X.-L. Liu, Y.-F. Gao, and J. Zhang, Probing the Magnetic Ordering of Antiferromagnetic MnPS3 by Raman Spectroscopy, J. Phys. Chem. Lett. **10**, 3087 (2019).
[77] A. V. Peschanskii, T. Y. Babuka, K. E. Glukhov, M. Makowska-Janusik, S. L. Gnatchenko, and Y. M. Vysochanskii, Raman study of a magnetic phase transition in the MnPS3 single crystal, Low. Temp. Phys. **45**, 1082 (2019).